\documentclass{emulateapj}

\shorttitle{Constraints on $f_{nl}$ for the WMAP data}
\shortauthors{Curto et al.}

\usepackage{graphicx}
\usepackage{float}
\usepackage{amsmath}
\usepackage{rotating}
\usepackage{subfigure}
\usepackage{epsfig,floatflt}

\begin{document}
%
\title{Improved constraints on primordial non-Gaussianity for \\ the Wilkinson Microwave Anisotropy Probe  5-yr data}
\author{A.\ Curto\altaffilmark{1}, E.\ Mart\'{\i}nez-Gonz\'alez and
  R. B.\ Barreiro}\affil{IFCA, CSIC-Univ. de Cantabria, Avda. los
  Castros, s/n, E-39005-Santander, Spain}

\altaffiltext{1}{Also at Dpto. de F\'{\i}sica 
Moderna, Univ. de Cantabria, Avda. los Castros, s/n, 39005-Santander, Spain}
\email{curto@ifca.unican.es}
%
%
%
%
%
%
%
\begin{abstract}
We present constraints on the non-linear coupling parameter $f_{nl}$
with the Wilkinson Microwave Anisotropy Probe (WMAP) data.  We use an
updated method based on the spherical Mexican hat wavelet (SMHW) which
provides improved constraints on the $f_{nl}$ parameter. This paper is
a continuation of a previous work by Curto et al. where several third
order statistics based on the SMHW were considered. In this paper, we
use all the possible third order statistics computed from the wavelet
coefficient maps evaluated at 12 angular scales. The scales are
logarithmically distributed from 6.9 arcmin to 500 arcmin. Our
analysis indicates that $f_{nl}$ is constrained to $-18 < f_{nl} <
+80$ at 95\% confidence level (CL) for the combined V+W WMAP map. This
value has been corrected by the presence of undetected point sources,
which adds a positive contribution of $\Delta f_{nl} = 6 \pm 5$. Our
result excludes at $\sim$99\% CL the best-fitting value $f_{nl}=$87
reported by Yadav \& Wandelt. We have also constrained $f_{nl}$ for
the Q, V and W frequency bands separately, finding compatibility with
zero at 95 \% CL for the Q and V bands but not for the W band. We have
performed some further tests to understand the cause of this deviation
which indicate that systematics associated to the W radiometers could
be responsible for this result. Finally we have performed a Galactic
North-South analysis for $f_{nl}$. We have not found any asymmetry,
i.e. the best-fitting $f_{nl}$ for the northern pixels is compatible
with the best-fitting $f_{nl}$ for the southern pixels.
\end{abstract}

\section{Introduction}
The cosmic microwave background (CMB) offers a picture of the early
Universe when it was only 400,000 years old. The CMB photons last
scattered off electrons at that time and since then they traveled free
through the space. The primordial perturbations set up during
inflation are imprinted in both radiation and matter distribution. The
CMB temperature anisotropies, related with the primordial
perturbations, can be used to test some assumptions of the so called
standard model. In particular, we can test the prediction of the
standard, single field, slow roll inflation
\citep{guth,albrecht,linde1982,linde1983} which states that the
anisotropies are compatible with a nearly Gaussian random field. Other
non-standard models of inflation predict detectable levels of
non-Gaussianity in the anisotropies \citep[see e.g.][]{bartolo}.

Primordial non-Gaussianity of the local form is characterised by the
non-linear coupling parameter $f_{nl}$
\citep{salopek1990,gangui1994,verde2000,komatsu2001}:
\begin{equation}
\Phi({\bf x}) = \Phi_L({\bf x}) + f_{nl}\{\Phi_L^2({\bf x})-\langle \Phi_L^2({\bf x}) \rangle \} \ \ 
\end{equation}
where $\Phi({\bf x})$ is the primordial gravitational potential and
$\Phi_L({\bf x})$ is a linear random field which is Gaussian
distributed and has zero mean.

There are many studies based on different statistical tools to
constrain the local $f_{nl}$ parameter from the CMB anisotropies,
using the data of different experiments. We can mention the analyses
using the angular bispectrum and wavelets on the Cosmic Background
Explorer (COBE) data \citep{komatsu2002,cayon2003}, the angular
bispectrum on MAXIMA data \citep{santos2003}, the angular bispectrum
on WMAP data
\citep{komatsu2003,creminelli2006,creminelli2007,spergel2007,yadav2008,komatsu2008,smith2009},
different kind of wavelet analyses on WMAP data
\citep{mukherjee2004,cabella2005,curto2008a}, the Minkowski
functionals on BOOMERanG data \citep{troia2007}, the Minkowski
functionals on Archeops data \citep{curto2007,curto2008}, the
Minkowski functionals on WMAP data
\citep{komatsu2003,spergel2007,gott2007,hikage2008,komatsu2008} among
others. We can also mention new promising techniques as for example
one based on the $n$-point probability density distribution
\citep{vielva2008}, and other based on needlets
\citep{pietrobon2008,rudjord2009}. Other works are based on the use of
the large scale structure to constrain $f_{nl}$ \citep{slosar2008}.

This work is a continuation of the wavelet-based analysis by
\citet{curto2008a} of the WMAP\footnote{http://map.gsfc.nasa.gov/}
data. We use high resolution WMAP data maps, and compute the wavelet
coefficients for 12 angular scales logarithmically spaced from 6.9
arcmin to 500 arcmin. With these wavelet coefficients we compute all
the possible third order moments involving these scales.

The article is organised as follows. Section \ref{method} presents the
estimators used to test Gaussianity and to constrain $f_{nl}$, the
data maps and the simulations. Section \ref{results} summarises the
main results of this work and the conclusions are in Section
\ref{conclusions}.
\section{Methodology}
\label{method}
This analysis is based on the spherical Mexican hat wavelet (SMHW) as
defined in \citet{martinez2002}. For references about the use of the
SMHW to test Gaussianity in the CMB see for example the review by
\citet{martinez2008}. We compute the wavelet coefficient maps at
several scales $R_i$ logarithmically separated ($R_{i+1}/R_i$
constant). The considered scales are\footnote{The level of
  discretisation of the wavelet space is a balance between the minimum
  number of scales needed to extract the non-Gaussian signal from the
  data and acceptable computational requirements.}: $R_1=6.9'$,
$R_2=10.6'$, $R_3=16.3'$, $R_4=24.9'$, $R_5=38.3'$, $R_6=58.7'$,
$R_7=90.1'$, $R_8=138.3'$, $R_9=212.3'$, $R_{10}=325.8'$,
$R_{11}=500'$. We also include the unconvolved map, which will be
represented by the scale $R_0$ as in \citet{curto2008a}. For each
possible combination of three scales $R_i$, $R_j$, and $R_k$ (where
the indices $i$, $j$, and $k$ can be repeated) we define a third order
statistic
\begin{equation}
\label{statistic}
q_{i j k}=\frac{1}{N_{i,j,k}}\sum_{p=0}^{N_{pix}-1}\frac{w_{p,i}w_{p,j}w_{p,k}}{\sigma_i\sigma_j\sigma_k}
\end{equation}
where $N_{pix}$ is the total number of pixels of the map, $N_{i,j,k}$
is the number of pixels available after combining the extended
  masks corresponding to the three scales $R_i$, $R_j$ and $R_k$,
$w_{p,i}=w_p(R_i)$ is the wavelet coefficient in the pixel $p$
evaluated at the scale $R_i$, and $\sigma_i$ is the dispersion of
$w_{p,i}$. Each map $w_{p,i}$ is masked out with the corresponding
extended mask at the scale $R_i$ as in \citet{curto2008a}. For a set
of $n$ scales we have $n_{stat}=(n+3-1)!/[3!(n-1)!]$ third order
statistics such as the one defined in Eq. (\ref{statistic}). We
  have tested with simulations that these statistics have a
Gaussian-like distribution. We can construct a vector ${\bf q}$ of
dimension $n_{stat}$
\begin{equation}
\label{vector}
{\bf q} = [q_{0,0,0}; q_{0,0,1}; ...; q_{0,0,11}; q_{0,1,1}; ...; q_{11,11,11}].
\end{equation}
With this vector we can perform two different analyses using a
$\chi^2$ statistic: one to test Gaussianity and a second one to
constrain $f_{nl}$ \citep[Eqs. (7) and (8) of][]{curto2008a}.

We use the 5-yr WMAP foreground reduced data, available in the Legacy
Archive for Microwave Background Data Analysis (LAMBDA) web
site\footnote{http://lambda.gsfc.nasa.gov}. We combine the maps of
different radiometers using the inverse of the noise variance as an
optimal weight \citep{bennett2003}. In particular we analyse the V+W,
Q, V and W combined maps at a resolution of 6.9 arcmin, corresponding
to a HEALPix \citep{healpix} $N_{side}=512$. We use the $KQ$75 mask
and also a set of extended masks for the wavelet coefficient maps. We
use the same masks as the ones described in \citet{curto2008a} for a
threshold of 0.01. This corresponds to an available fraction of
the sky from 71.2\% for the $R_1$ scale to 31.4\% for the $R_{11}$
scale. Notice that larger scales have the restriction of a lower
available area, which means a lower sensitivity to $f_{nl}$.

Finally we analyse the data and compare them with Gaussian and
non-Gaussian simulations. The Gaussian simulations are performed using
the best fit power spectrum $C_{\ell}$ for WMAP provided by LAMBDA and
the instrumental white noise of each WMAP radiometer. The non-Gaussian
simulations with the $f_{nl}$ contribution are computed following the
algorithms described in \citet{liguori2003,liguori2007} and
transformed into WMAP maps with the instrumental noise included. We
also estimate the unresolved point source contribution to $f_{nl}$ for
the V+W case by analysing point source simulations. These simulations
have been generated as in \citet{curto2008a} following the source
number counts $dN/dS$ given by \citet{zotti2005}.
\section{Results}
\label{results}
In this Section we present the Gaussianity analysis of the WMAP data
for the combined V+W, Q, V, and W data maps. We also constrain
$f_{nl}$ for these maps through a $\chi^2$ test. We estimate the
contribution of point sources to the V+W map. Finally we constrain
$f_{nl}$ for northern and southern pixels separately.
\subsection{Analysis of WMAP data}
\begin{table}
  \center
  \caption{$\chi^2$ constructed from the 364 statistics for V+W, Q, V,
    and W data maps. \label{stats_vw} We also present the mean and the
    dispersion of the $\chi^2$ corresponding to 1,000 Gaussian
    simulations, and the cumulative probability for the $\chi^2$ of
    the data obtained from the simulations.}
  \begin{tabular}{cccccc}
    \hline 
    \hline
    Map & $\chi_{data}^2$ & DOF & $\langle \chi^2 \rangle$ & $\sigma$ & $P(\chi^2\le \chi_{data}^2)$\\
    \hline
      V+W & 349 & 364 & 379 & 49.2 & 0.29 \\
      Q & 384 & 364 & 378 & 48.7 & 0.63 \\
      V & 348 & 364 & 377 & 47.8 & 0.27 \\
      W & 354 & 364 & 376 & 47.9 & 0.35 \\
    \hline
    \hline
  \end{tabular}
\flushleft
\end{table}
\begin{figure*}
\center
\includegraphics[width=8.5cm,height=5.3cm] {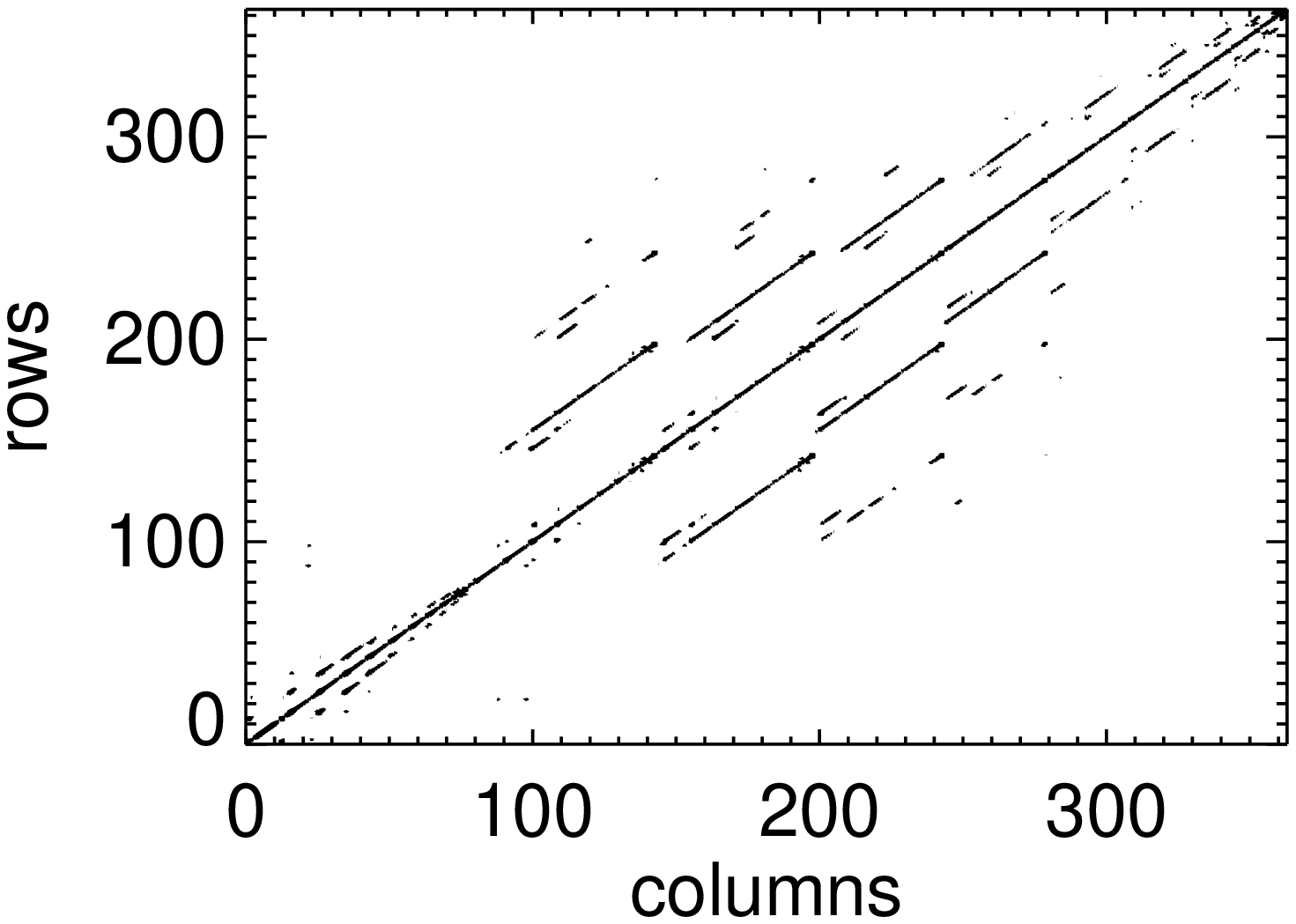}
\includegraphics[width=8.5cm,height=5.3cm] {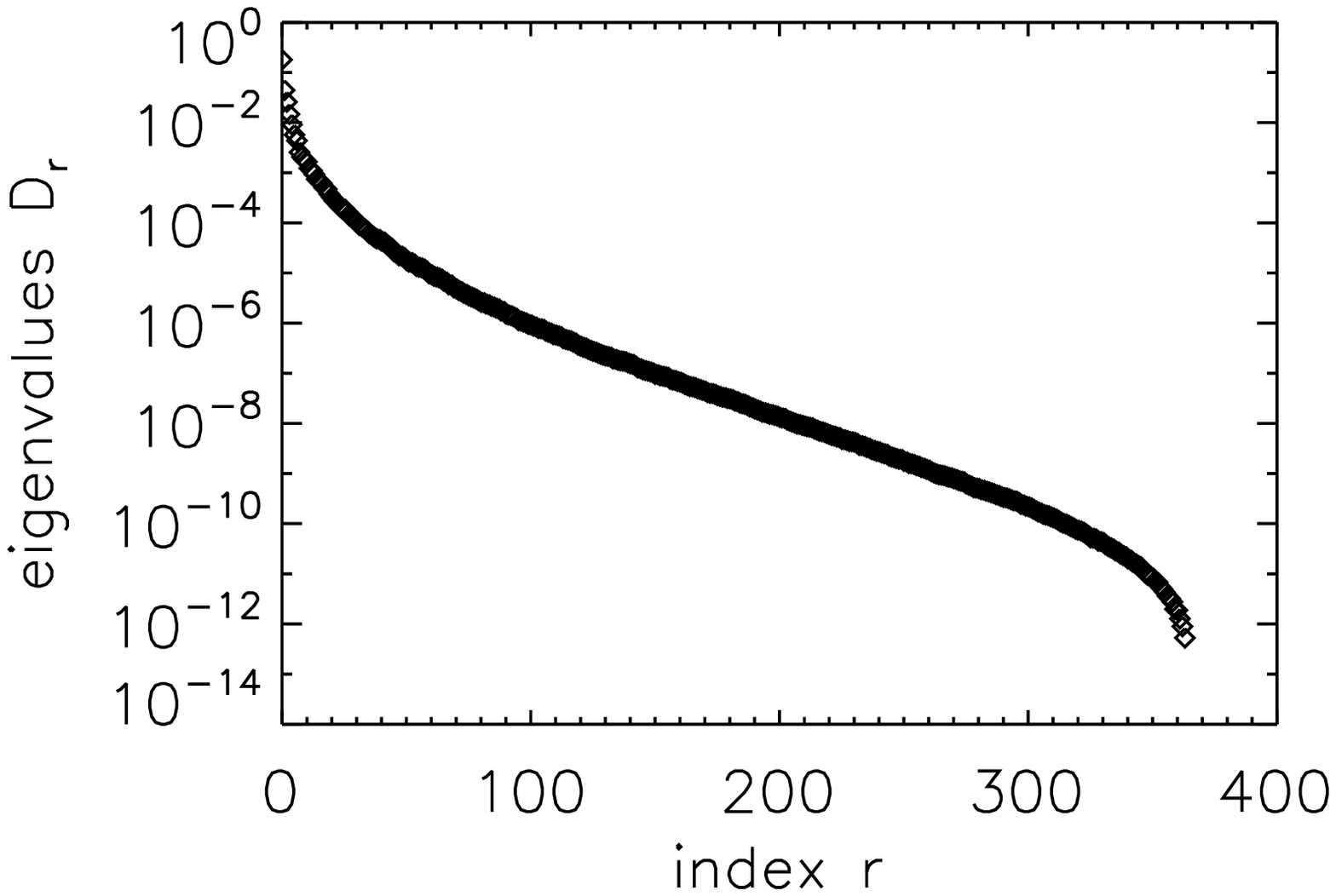}
\caption{Left panel: the covariance matrix elements of the third order
  statistics for the V+W map that satisfy
  $C_{ijk,rst}/\sqrt{C_{ijk,ijk}C_{rst,rst}} >$ 0.95. Right panel: the
  covariance matrix eigenvalues for the V+W map.}
\label{covariance_matrix_properties}
\end{figure*}
\begin{figure}
\center
\includegraphics[width=8.5cm,height=5.3cm] {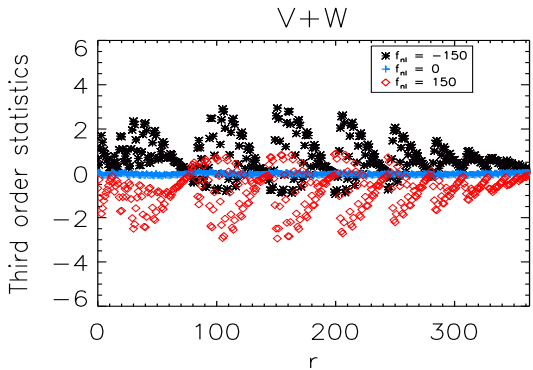}
\caption{The expected values of the normalised third order statistics
  $q_r$ for V+W simulations with different values of $f_{nl}$,
    where $r \equiv \{i,j,k\}$ is ordered following
    Eq. (\ref{vector}).}
\label{trd_stat_vs_fnl}
\end{figure}
\begin{figure}
\center
\includegraphics[width=8.5cm,height=5.3cm] {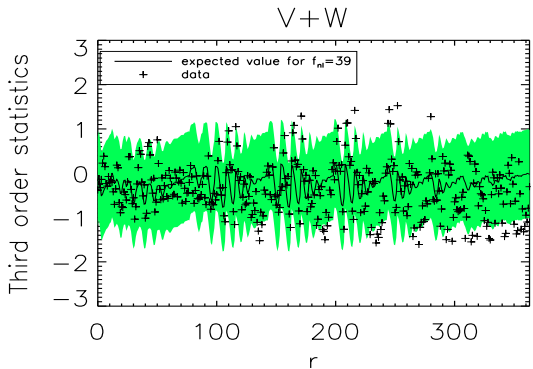}
\caption{The normalised third order statistics $q_r$ for the combined
  V+W data map. The solid line corresponds to the expected values for
  the best-fitting $f_{nl}$ model ($f_{nl}=39$). We also plot the
  1$\sigma$ error bars in green.}
\label{data_trd_stat_vs_fnl}
\end{figure}
\begin{table}
  \center
  \caption{Best-fitting $f_{nl}$ values obtained from V+W, Q, V
    and W combined maps. We also present the mean, dispersion and some
    percentiles of the distribution of the best fit $f_{nl}$ values
    obtained from Gaussian simulations. \label{bestfnl_vw}}
  \begin{tabular}{@{}c@{~~}c@{~~}c@{~~}c@{~~}c@{~~~}c@{~~~}c@{~~~}c@{}}
    \hline 
    \hline
    Map & best $f_{nl}$ & $\langle f_{nl} \rangle$ & $\sigma(f_{nl})$ &  $X_{0.160}$ & $X_{0.840}$ & $X_{0.025}$ & $X_{0.975}$ \\
    \hline
      V+W & 39 & -1 & 25 & -26 & 24 & -51 & 47 \\
      Q & 11 & 0 & 33 & -31 & 34 & -63 & 66 \\
      V & 23 & 0 & 30 & -28 & 30 & -55 & 59 \\
      W & 65 & -4 & 30 & -33 & 26 & -59 & 58 \\
    \hline
    \hline
  \end{tabular}
\flushleft
\end{table}
\begin{figure*}
\center
\includegraphics[width=8.5cm,height=5.3cm] {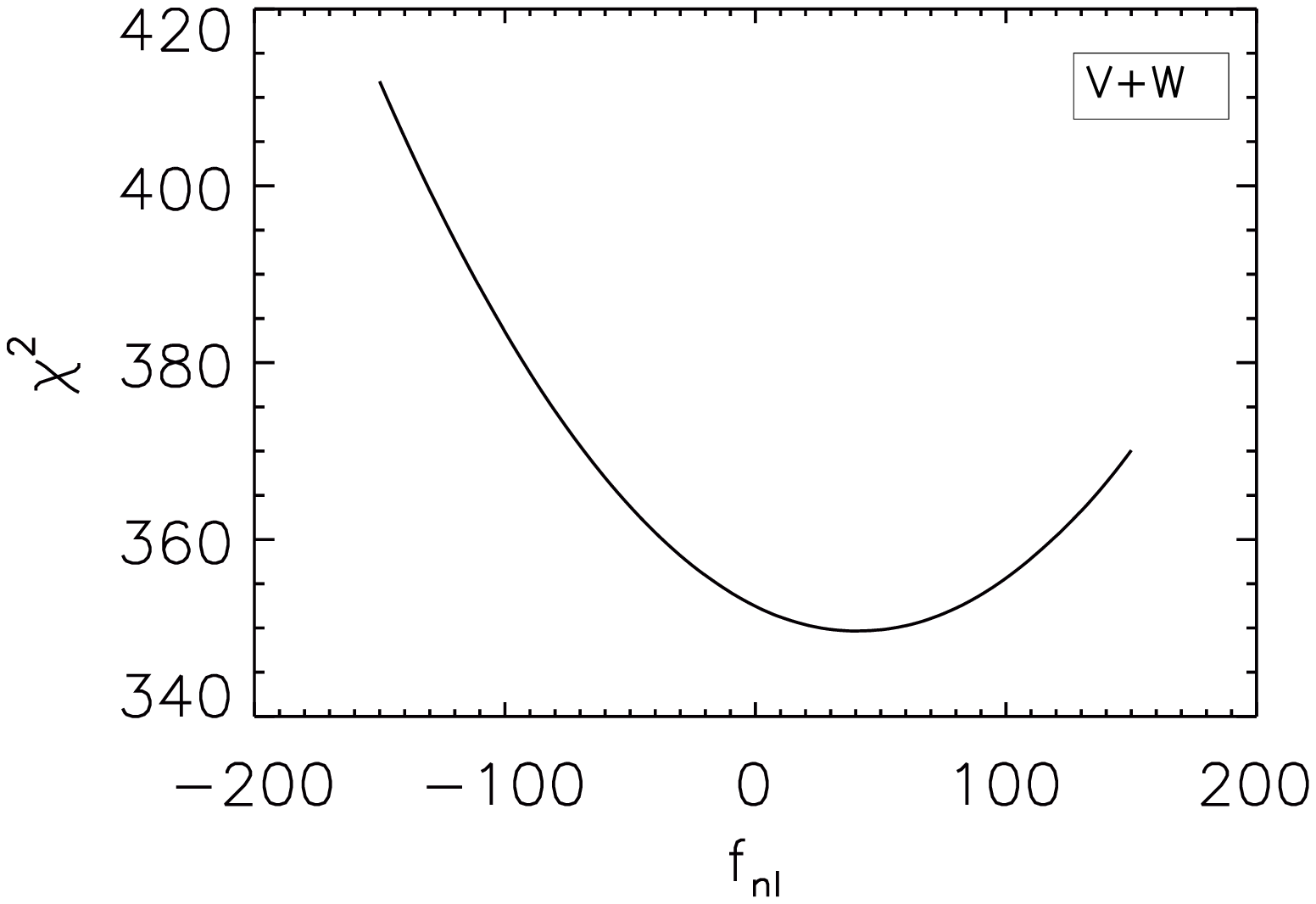}
\includegraphics[width=8.5cm,height=5.3cm] {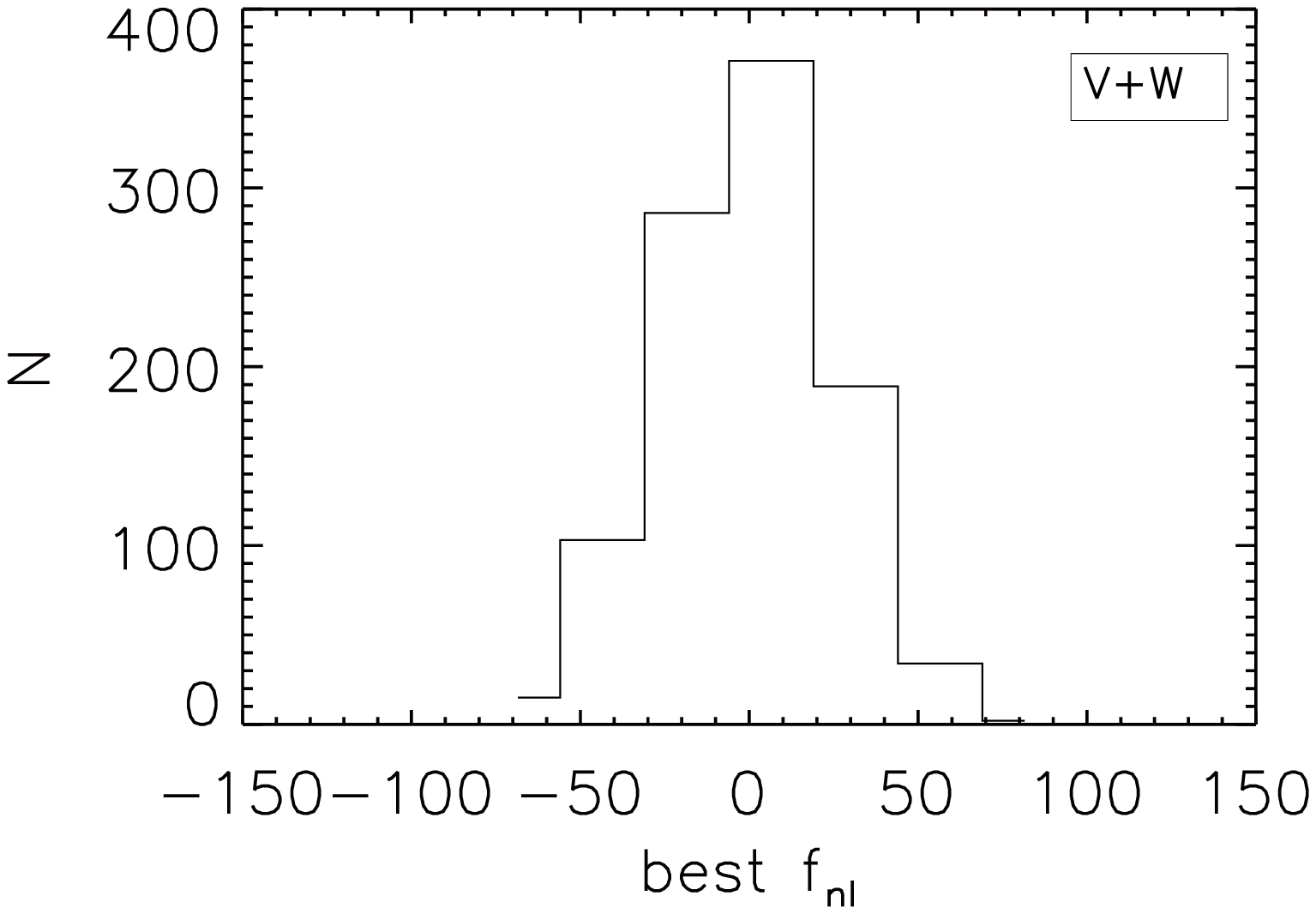}
\caption{The $\chi^2(f_{nl})$ statistics versus $f_{nl}$ for the V+W
  data map and the histogram for the best-fitting $f_{nl}$ values for
  a set of 1000 Gaussian V+W simulations. The dispersion is
  $\sigma(f_{nl})=25$.}
\label{bestfnl_data_sims_vw}
\end{figure*}
We evaluate the wavelet coefficients at the 12 considered scales. With
these coefficients we compute the third order estimators defined in
Eq. (\ref{statistic}). For 12 scales, we have 364 possible third order
statistics. We compute these statistics for the data maps and for
Gaussian simulations. The covariance matrix used in the $\chi^2$
statistics is constructed from 10,000 Gaussian simulations. For the
considered cases, V+W, Q, V and W, we have that the data are inside
the $2\sigma$ error bars, i.e., the data are compatible with Gaussian
simulations. We compute a $\chi^2$ statistic by comparing the data
with the expected value of Gaussian simulations following
\citet{curto2008a}. We as well compute the $\chi^2$ of an additional
set of 1,000 Gaussian simulations. The $\chi^2$ statistic of the data
is compatible with the $\chi^2$ of the Gaussian simulations. This is
presented in Table \ref{stats_vw}. 

We have studied the properties of the covariance matrix for the V+W
map (see left panel of Fig. \ref{covariance_matrix_properties}) which
is derived from 10,000 Gaussian simulations. We may wonder if that
number of simulations is enough to achieve the convergence. The
condition number, defined here as the ratio between the maximum and
minimum eigenvalues (see right panel of
Fig. \ref{covariance_matrix_properties}) is $Cond(C) \sim 10^{12}$. An
upper limit for the relative errors in the inverse covariance matrix
is $Cond(C) \times \epsilon_{mach}$, where $\epsilon_{mach}$ is the
computer error, $\epsilon_{mach} \sim 10^{-16}$ for double
precision. Thus the relative error for the inverse matrix coefficients
is $Cond(C) \times \epsilon_{mach} \sim 10^{-4}$ which is almost
negligible. We have also computed the covariance matrix with two
independent sets of 5,000 simulations and constrained the best-fitting
$f_{nl}$ of the data with these two matrices. The best-fitting
$f_{nl}$ value of the data is almost the same for both matrices and it
agrees with the result obtained with 10,000 simulations suggesting
that the convergence is almost reached with 5,000 simulations. 

We impose constraints on $f_{nl}$ through a $\chi^2$ test. We
calculate the expected values of the estimators for different $f_{nl}$
cases using a set of 300 non-Gaussian simulations. In
Fig. \ref{trd_stat_vs_fnl} we plot the expected values of the 364
statistics $q_r$ for several $f_{nl}$ cases for the V+W map. We have
checked that the estimator is unbiased. We have evaluated the expected
values of the estimator using 200 non-Gaussian simulations of the V+W
map and analysed the remaining 100 independent non-Gaussian
simulations with $f_{nl}=50$.  The result is an average
best-fitting $f_{nl}$ value of 50.6.

In Fig. \ref{data_trd_stat_vs_fnl} we plot the values of the
statistics for the data map and compare them with the expected values
for the best-fitting $f_{nl}$ model.  Then we perform a $\chi^2$
analysis to find the best-fitting value of $f_{nl}$ for each map. We
also analyse Gaussian simulations in order to obtain the frequentist
error bars. Table \ref{bestfnl_vw} lists the best-fitting $f_{nl}$
values for the V+W, Q, V and W combined maps and the main properties
of the histograms of the best-fitting $f_{nl}$ obtained from Gaussian
simulations. In the left panel of Fig. \ref{bestfnl_data_sims_vw} we
plot the $\chi^2(f_{nl})$ versus $f_{nl}$ for the V+W map, and in the
right panel of Fig. \ref{bestfnl_data_sims_vw} we plot the histogram
of the best-fitting $f_{nl}$ of 1,000 Gaussian simulations.  We have
$-12 < f_{nl} < +86$ for V+W, $-52 < f_{nl} < +77$ for Q, $-32 <
f_{nl} < +82$ for V and $+6 < f_{nl} < +123$ for W (all at 95\%
CL). Note that $f_{nl}$ increases as the frequency grows from Q to V
and W bands. This suggests the possible presence of foregrounds
residuals as they are more important at low frequencies and they add a
negative contribution to $f_{nl}$ for the bispectrum estimator
\citep{yadav2008}. To further check if this is also the case for the
wavelet estimator, we have also studied the $K$ and $Ka$ bands, where
the foreground signal has an important contribution. We have obtained
$f_{nl}^K =-497 \pm 42$ and $f_{nl}^{Ka} =-18 \pm 37$. Taking this
into account and the values of $f_{nl}$ for the clean and raw Q, V,
and W maps (see Tables \ref{bestfnl_vw} and \ref{bestfnl_w_band}) we
can see that our estimator is sensitive to the presence of foregrounds
(biasing the result towards lower values). The sensitivity is even
more significant for the Q, V, and W bands using the bispectrum
\citep[see Table 6 in][]{komatsu2008}.

It is interesting to point out that the wavelet-based method has an
intermediate dimension ($n_{stat} = $ 364 for 12 scales) when compared
with the bispectrum combinations ($\ell_{max}^{\beta}$ where $2 <
\beta < 3$ and $\ell_{max} \sim 10^3$) and the Minkowski
functionals (usually several tens of combinations). The number of
wavelet cubic combinations increases significantly with the number of
scales.

We also estimate the contribution of the point sources for the V+W
combined map as in \citet{curto2008a}. We add the point source
simulations to the CMB plus noise simulations. For each one of them we
compute its best-fitting $f_{nl}$ and compare it with the obtained for
the same case without including the point source simulation. The
difference returns an estimate of the contamination on $f_{nl}$ due to
point sources. For the V+W map we have $\Delta f_{nl} = 6 \pm
5$. Therefore our estimate taking into account the point sources is
$-18 < f_{nl} < +80$ at 95\% CL for the V+W map. Comparing with the
best-fitting value for the V+W map given by \citet{yadav2008},
$f_{nl}=87$, our analysis excludes that value at $\sim$99\% CL.
However, it is important to point out that \citet{yadav2008} used
different choices for the data maps in the analysis (they used WMAP
3-yr data whereas we use WMAP 5-yr data, different weighting for the
channels, etc.), which could also contribute to the found
discrepancy. In particular, a simple average of the channels (instead
of a noise weighted combination) enhances the signal at high
multipoles at the cost of having noisier maps.
\subsection{Constraints on $f_{nl}$ with the W band}
The W map best-fitting $f_{nl}$ value is only compatible with zero
at 99\% CL. This result is in apparent discrepancy with the values
obtained for the V and V+W maps, which are compatible with zero at
95\%CL. The point sources add a low contribution to the W map,
$\Delta f_{nl}=1\pm2$, and therefore they do not explain its
best-fitting value of $f_{nl}=$65.  We may wonder if that value can be
obtained by a statistical fluctuation.  Considering simulations with
different models ($f_{nl}=0$, $f_{nl}=40$ and $f_{nl}=70$) we have
confirmed that the best-fitting $f_{nl}$ for W is compatible with
the values obtained for the V and V+W maps.  

To understand the relatively large $f_{nl}$ value found in the W map
we have performed some additional tests.  First of all, we have
checked if this deviation could be due to the presence of residual
foregrounds by studying the V-W map for the clean and the raw (before
template subtraction) maps as well as the raw maps for the Q,V,W and
V+W cases (see Table \ref{bestfnl_w_band}). We find that there is a
very significant (positive) deviation in the clean V-W map.  Notice
that the V-W map has CMB residuals due to the different resolutions of
the V and W bands, which have also been taken into account in the V-W
simulations. In any case, the signal due to the CMB is very small and
we do not expect it to affect the results given in Table
\ref{bestfnl_w_band} for the different maps. Since this combination
contains mainly residual foregrounds and noise, both could be
responsible for the deviation. Interestingly, when we repeat the test
for the raw V-W map, where foreground contamination should be more
important, the best $f_{nl}$ value becomes compatible with the
simulations. This indicates again that foreground emission tends to
bias the estimated $f_{nl}$ towards lower values. This is also
observed for the best $f_{nl}$ value estimated for the raw Q, V, W and
V+W maps, which is systematically lower than the one obtained for the
clean maps. Therefore, for the case of the raw V-W map, some effect
from systematics may be cancelled by foreground residuals.

If foregrounds are not responsible for the deviation found in the W
and V-W maps, we may wonder if it is due to systematics present in the
W radiometers. To test this possibility, we have studied two different
combinations of the four W radiometers, where CMB and foregrounds are
basically cancelled. One of these combinations is consistent with
Gaussian simulations but the second one shows again a deviation at the
95 per cent CL, indicating the possible presence of some spurious
signal in the noise of one or several of the W radiometers. In order
to localise further the origin of this signal, we have also studied
each W radiometer separately, finding a deviation at the level of 98
per cent for the $W_2$ radiometer, with a best $f_{nl}$ value of 91,
while the rest of the radiometers are consistent with the zero value
at the 95 per cent CL. Finally we have also considered the difference
between the two V radiometers, which is found to be compatible with
Gaussianity. This further indicates that, if a systematic is
responsible of the V-W map, this would be present in the W frequency
channel (see Table \ref{bestfnl_w_band}).

All these tests suggest that the relatively large $f_{nl}$ value
obtained for W may come from systematics present in the W radiometers.
In any case, a more exhaustive study is necessary in order to
establish the origin of this deviation.
\begin{table}
  \center
  \caption{Best-fitting $f_{nl}$ values for different WMAP radiometers
    and combinations of them for raw and clean data. We also present
    the mean and the dispersion of the best-fitting $f_{nl}$ values
    obtained from Gaussian simulations. \label{bestfnl_w_band}}
  \begin{tabular}{@{}c@{~~}c@{~~}c@{~~}c@{~~}c@{~~}c@{}}
    \hline 
    \hline
    Map & foreground & best $f_{nl}$ & $\langle f_{nl} \rangle$ & $\sigma(f_{nl})$ \\ 
    \hline
    V+W & raw & 34 & -1 & 25 \\
    Q & raw & -3 & 0 & 33 \\
    V & raw & 16 & 0 & 30 \\
    W & raw & 60 & 0 & 30 \\
    V-W & raw & -0.02 & 0.02 & 0.29 \\
    \hline
    V-W & clean & 1.02 & 0.02 & 0.29 \\ 
    $W_1$ & clean & 39 & 1 & 41 \\ 
    $W_2$ & clean & 91 & -3 & 45 \\ 
    $W_3$ & clean & 23 & 2 & 47 \\ 
    $W_4$ & clean & 59 & 0 & 44 \\ 
    $W_1+W_2-W_3-W_4$ & clean & -0.52 & 0.00 & 0.34 \\ 
    $W_1-W_2+W_3-W_4$ & clean & 0.85 & 0.00 & 0.35 \\ 
    $V_1-V_2$ & clean & -0.01 & 0.01 & 0.34 \\ 
    \hline
    \hline
  \end{tabular}
\flushleft
\end{table}
\subsection{$f_{nl}$ for the North and South hemispheres}
The localization property of the wavelets allows a local analysis of
the $f_{nl}$ parameter. In particular, we test the Gaussianity and
estimate the best-fitting $f_{nl}$ value of the V+W combined map
using only northern (Galactic latitude $b > 0$) and southern pixels
($b < 0$) in Eq. \ref{statistic}. For both cases the third order
statistics obtained for the data are compatible with Gaussian
simulations (inside the $2\sigma$ error bars). In Table
\ref{bestfnl_vw_hemispheres} we list the best-fitting $f_{nl}$ values
for the northern and southern hemispheres and the main properties of
the distribution of the best-fitting $f_{nl}$ obtained from 1,000
Gaussian simulations.  We estimated the contribution of the unresolved
point sources as in the previous subsection, and the values are
$\Delta f_{nl} = 7\pm7$ for the North and $\Delta f_{nl} = 5\pm7$ for
the South. Taking into account this, the results are $-32 < f_{nl} <
113$ for the North and $-50 < f_{nl} < 99$ for the South at 95\%
CL. These values are compatible with zero at 95\% CL. We also study
the compatibility of the best-fitting value for the North and South
hemispheres between them. We compute the difference of the
best-fitting $f_{nl}$ value $\Delta f_{nl}=f_{nl}^{(N)} -
f_{nl}^{(S)}$ for the North and South hemispheres for the set of 1,000
V+W Gaussian simulations. The difference is $\Delta f_{nl}^{(data)}
=11$ for the data, and for the simulations is $ \Delta f_{nl} =
3\pm55$. The cumulative probability is $ P(\Delta f_{nl} \le \Delta
f_{nl}^{(data)}) = 0.57$ and therefore the difference for the data is
compatible with the results obtained from simulations. This means that
we do not find any asymmetry in the North-South $f_{nl}$ value.
\begin{table}
  \center
  \caption{Best-fitting $f_{nl}$ values obtained for the northern and
    southern hemispheres. We also present the mean, dispersion and some
    percentiles of the distribution of the best fit $f_{nl}$ values
    obtained from Gaussian simulations. \label{bestfnl_vw_hemispheres}}
  \begin{tabular}{@{}c@{~~}c@{~~}c@{~~}c@{~~}c@{~~~}c@{~~~}c@{~~~}c@{}}
    \hline 
    \hline
    Region & best $f_{nl}$ & $\langle f_{nl} \rangle$ & $\sigma(f_{nl})$ &  $X_{0.160}$ & $X_{0.840}$ & $X_{0.025}$ & $X_{0.975}$ \\
    \hline
      North & 46 & 2 & 37 & -35 & 40 & -71 & 74 \\
      South & 35 & -1 & 38 & -39 & 37 & -80 & 69 \\
    \hline
    \hline
  \end{tabular}
\flushleft
\end{table}
\section{Conclusions}
\label{conclusions}
We have tested the Gaussianity and constrained the $f_{nl}$ parameter
with the 5-yr WMAP data. We use an optimal wavelet-based test. We have
considered the V+W, Q, V and W combined maps at high
resolution. We have used a set of 300 realistic non-Gaussian
simulations and thousands of Gaussian simulations for the
analysis. We have computed the wavelet coefficient maps at scales from
6.9 arcmin to 500 arcmin and computed all the possible third order
moments (Eq. \ref{statistic}) using appropriate extended masks.

The data are compatible with Gaussian simulations for the considered
combined maps (see Table \ref{stats_vw}). We have imposed constraints
on the non-linear coupling parameter $f_{nl}$ by using non-Gaussian
simulations with $f_{nl}$. The results show that $f_{nl}$ increases
when we go from the Q to the V and W combined maps. This frequency
dependence also appears in the results by
\citet{yadav2008,komatsu2008,curto2008a}. The results are compatible
with zero at 95\% CL for the V+W, Q, and V combined maps, but not for
the W map (which is compatible at 99\% CL). This value cannot be
explained by unresolved point sources since their contribution is
$\Delta f_{nl}=1\pm2$ for the W map. We have estimated the probability
of having those values with simulations and the results do not show
incompatibility among different channels.  We have also seen that the
relatively large $f_{nl}$ value obtained for the W band may come from
systematics in one or several radiometers of this band.

We have also estimated the contribution of unresolved point sources to
$f_{nl}$ for the V+W map using a realistic model given by
\citet{zotti2005}. The results are $\Delta f_{nl} = 6\pm5$. Taking
into account this value, our best estimate for $f_{nl}$ is $-18 <
f_{nl} < +80$ at 95\% CL. The use of new scales and all the possible
third order moments has returned better constrains to $f_{nl}$ and
lower error bars compared with the results by \citet{curto2008a} and
previous works. Our best estimate is compatible with the values
obtained by \citet{komatsu2008} and it excludes the best-fitting
$f_{nl}$ value obtained by \citet{yadav2008} at the $\sim$99\% CL.

Finally we have constrained $f_{nl}$ for the North and South
hemispheres and the results give two best-fitting values that are
compatible with zero at 95\% CL and also are compatible between
them. Therefore we do not find any North-South asymmetry for this
parameter.
\section*{acknowledgments}
We are thankful to Frode Hansen, Michele Liguori and Sabino Matarrese
for supplying maps with primordial non-Gaussianity.  The authors thank
J. Gonz\'alez-Nuevo for providing the $ dN/dS$ counts and for his
useful comments on unresolved point sources. The authors thank
P. Vielva, E. Komatsu and M. Cruz for useful discussions. We also
thank R. Marco and L. Cabellos for computational support. We
acknowledge partial financial support from the Spanish Ministerio de
Ciencia e Innovaci\'on project AYA2007-68058-C03-02. A. C. thanks the
Spanish Ministerio de Ciencia e Innovaci\'on for a pre-doctoral
fellowship.
The authors acknowledge the computer resources, technical expertise
and assistance provided by the Spanish Supercomputing Network (RES)
node at Universidad de Cantabria. We acknowledge the use of Legacy
Archive for Microwave Background Data Analysis (LAMBDA). Support for
it is provided by the NASA Office of Space Science. The HEALPix
package was used throughout the data analysis \citep{healpix}.
%
%

%
\end{document}